\documentstyle[]{aa} 

\newcommand{\gpc}{\mbox{GP~Com}} 
\newcommand{\etal}{\mbox{et\ al.\ }}
\newcommand{\kmsec}{\,\mbox{$\mbox{km}\,\mbox{s}^{-1}$}}

\newcommand{\hea}{\hbox{$\hbox{He\,{\sc i}\,$\lambda$4388\,\AA}$}}
\newcommand{\heb}{\hbox{$\hbox{He\,{\sc i}\,$\lambda$4472\,\AA}$}}
\newcommand{\hec}{\hbox{$\hbox{He\,{\sc i}\,$\lambda$4713\,\AA}$}}
\newcommand{\hed}{\hbox{$\hbox{He\,{\sc i}\,$\lambda$4922\,\AA}$}}
\newcommand{\hee}{\hbox{$\hbox{He\,{\sc i}\,$\lambda$5016\,\AA}$}}
\newcommand{\hef}{\hbox{$\hbox{He\,{\sc i}\,$\lambda$6678\,\AA}$}}
\newcommand{\heg}{\hbox{$\hbox{He\,{\sc i}\,$\lambda$7065\,\AA}$}}
\newcommand{\heh}{\hbox{$\hbox{He\,{\sc i}\,$\lambda$7281\,\AA}$}}
\newcommand{\heii}{\hbox{$\hbox{He\,{\sc ii}\,$\lambda$4686\,\AA}$}}

\newcommand{\nva}{\hbox{$\hbox{N\,{\sc v}\,$\lambda$1239\,\AA}$}}
\newcommand{\nvb}{\hbox{$\hbox{N\,{\sc v}\,$\lambda$1243\,\AA}$}}

\title{New results on \gpc}
\author{L.\ Morales-Rueda\inst{1} \and T.\, R.\ Marsh \inst{1} \and D.\,Steeghs
  \inst{1,2} \and E.\,Unda-Sanzana \inst{1} \and Janet\, H.\ Wood
  \inst{3} \and R.\, C.\ North \inst{1,4}} 
\institute{Department of Physics and Astronomy, Southampton
  University, Southampton SO17 1BJ, UK\\
  \email{lmr@astro.soton.ac.uk, trm@astro.soton.ac.uk} \and
  Harvard-Smithsonian Center for Astrophysics, 60 Garden Street,
  Cambridge, MA 02318, USA\\ \email{dsteeghs@head-cfa.harvard.edu}
  \and Astrophysics Group, School of Chemistry and Physics, Keele
  University, Staffordshire, ST5 5GB, UK \and Met Office, London Road,
  Bracknell, Berkshire, RG12 2SZ, UK } 

\date{Received ...; accepted
  ...}

\begin{document}

\abstract{
  
  We present high resolution optical and UV spectra of the 46\,min
  orbital period, helium binary, \gpc. Our data contains simultaneous
  photometric correction which confirms the flaring behaviour observed
  in previous optical and UV data. In this system all lines show a
  triple peaked structure where the outer two peaks are associated
  with the accretion disc around the compact object. The main aim of
  this paper is to constrain the origin of the central peak, also
  called ``central spike''. We find that the central spike contributes
  to the flare spectra indicating that its origin is probably the
  compact object. We also detect that the central spike moves with
  orbital phase following an S-wave pattern. The radial velocity
  semiamplitude of the S-wave is $\sim$10\,km s$^{-1}$ indicating that
  its origin is near the centre of mass of the system, which in this
  case lies very close to the white dwarf.  Our resolution is higher
  than that of previous data which allows us to resolve structure in
  the central peak of the line. The central spike in three of the
  He\,{\sc i} lines shows another peak blueshifted with respect to the
  main peak. We propose that one of the peaks is a neutral helium
  forbidden transition excited in a high electron density region. This
  forbidden transition is associated with the permitted one (the
  stronger peak in two of the lines). The presence of a high electron
  density region again favours the white dwarf as their origin.  We
  compute Doppler maps for the emission lines which show three
  emission regions: an accretion disc, a bright spot and an
  unidentified low velocity emission region associated with the
  central spike. We obtain modulation Doppler tomograms for some of
  the emission lines that map the anisotropic emission from the bright
  spot region. The HST UV spectra also show a strong flare component
  and confirm the under abundance of silicon in \gpc.

}
\maketitle

\section{Introduction}

The AM~CVn stars are close binaries in which a white dwarf accretes
from the stripped-down core of a giant star or a helium white dwarf
(Tutukov \&\ Federova 1989). Three possible channels have been
discussed in the literature regarding the formation of AM~CVn systems:
(a) a double degenerate channel (Tutukov \& Yungelson 1979), (b) a
channel in which the donor is a semi-degenerate star (Savonije et al.
1986) and (c) a channel in which the donor starts as a non-degenerate
post main sequence star (Podsiadlowski et al. in preparation). Near
relatives of the cataclysmic variables (CVs), the AM~CVn stars have
periods ranging from 5 to 65 min, and accretion discs composed of
$>99$ per cent helium by number, giving us a unique opportunity to
test the influence of composition upon disc physics.  Although only 8
or 9 of these systems are known, they are predicted in some studies to
have a space density about a factor of 2 higher than that of the CVs
of which some 700 are known (Tutukov \& Yungelson 1996).  Other
studies predict an observable population that agrees with the observed
one (Nelemans et al. 2001).  AM~CVn systems may be significant sites
of accretion-induced collapse producing either neutron stars or Type
Ia supernovae, a possibility denied to the CVs because of erosion of
their white dwarfs in nova explosions (Tutukov \& Yungelson 1996).

The theory of binary star evolution predicts that most AM~CVn systems
have low accretion rates (Tutukov \& Yungelson 1996). The disc
instability model (Osaki 1974) predicts that at low accretion rates,
discs can exist in a steady-state. Moreover, the threshold below which
the discs are steady is higher for helium than hydrogen discs, a
consequence of the higher ionisation energy of helium (Tsugawa \&\ 
Osaki 1997). As a result most AM~CVn systems should be faint, and may
fail to outburst, making them hard to spot.  Of the known systems,
\gpc\ (=G61-29), which has an orbital period of 46 min, is
intrinsically faint ($M_V > 11$) and has never shown outbursts, is
closest to the properties expected of the bulk of AM~CVns (Tutukov \&
Yungelson 1996). \gpc 's components are thought to be a
$\sim$0.5\,M$_{\odot}$ CO white dwarf accreting from a
0.02\,M$_{\odot}$ helium degenerate (Nather et al.  1981).

\begin{figure*}
\begin{picture}(100,0)(-270,250)
\put(0,0){\includegraphics{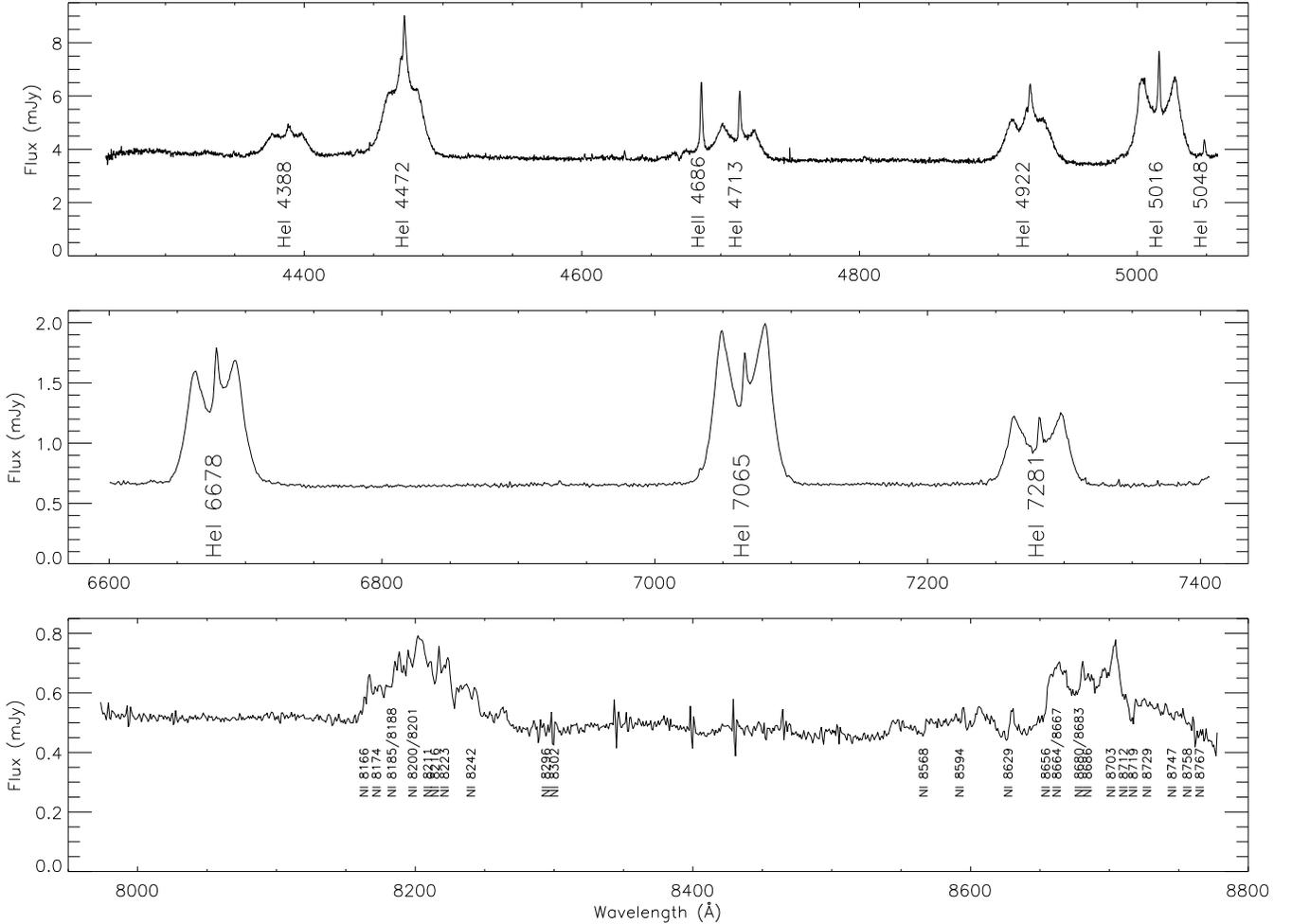}}
\end{picture}
\vspace{130mm}
\caption{Average spectra of all our \gpc\ data for the three wavelength ranges
  observed. The most prominent lines are labelled.}
\label{res:avspall}
\end{figure*}

\gpc 's optical spectrum, which is full of He\,{\sc i} and He\,{\sc
  ii} lines, resembles very closely that of a recently discovered
AM~CVn system, CE315 (Ruiz et al. 2001); the orbital period of CE315
is 65 minutes.  In the UV wavelength range, \gpc\ shows a plethora of
nitrogen lines and possibly some C\,{\sc i} but no sign of the C\,{\sc
  iv} or Si\,{\sc iv} commonly seen in CVs. Lambert \&\ Slovak (1981)
suggest that this is the result of the CNO process depleting the
carbon and enhancing the nitrogen.  Marsh et al. (1991) model the
optical lines with an LTE model and obtain a reasonable fit, which
confirms the CNO processing, but with very peculiar abundances of
heavy elements such as silicon, calcium and iron which appear to be
under abundant compared to the Sun by 100 to 1000 times.

\gpc\ flares erratically at X-ray (van Teeseling \&\ Verbunt 1994), UV
(Marsh et al. 1995) and optical wavelengths (Marsh et al. 1991). The
variability is strongest in high ionisation lines, implicating the
inner disc, as do the short-timescales involved. This suggests that
the flaring is driven by irradiation of the disc by a variable source
at the centre of the disc. The flaring causes large changes in the
line profiles (Marsh \etal 1995).  The line profiles show evidence for
being ionisation-bounded (Marsh 1999).

In this paper we study the source of the flaring and in particular the
contribution of the mystery spike at the centre of each line to the
line flaring. The origin of the central spike has been suggested to be
a nebula (Nather \etal 1981), but none has been found (Stover 1983).
Instead we see that the spike may participate in flaring (Marsh 1999),
which unlikely though it seems, leaves the accreting white dwarf as
its most plausible origin. A potential problem of Marsh's (1999) study
was that the spectra were only approximately photometric, with
corrections based upon a comparison star on the slit. This could have
made the central spike appear to be varying artificially.

We obtain modulation Doppler tomograms that map the flux variability
seen in the bright-spot region. We also present HST UV data and study
the flaring component in the UV spectra.

\section{Observations}

We obtained spectroscopy of \gpc\ on the nights of 1999 March 26 and
27, using the ISIS double beam spectrograph mounted on the 4.2\,m
William Herschel Telescope (WHT) on La Palma. Table~\ref{obs:jo} gives
a journal of observations. 

For the red arm, the standard readout mode was used in conjunction
with a Tektronix CCD(1k$\times$1k) windowed to 1124\,$\times$\,650
pixels to reduce dead time. Exposure times were generally 180\,s, for
the first night and 300\,s during the second night (when the seeing
got worse) with $\sim$\,33\,s dead time, and the spectral resolution
corresponds to $56\,\kmsec$ at $\lambda$7067\AA. The 600
lines\,mm$^{-1}$ grating R600R was used to cover the wavelength range
$\lambda\lambda$6600 -- 7408\AA\ during the first night and
$\lambda\lambda$7972 -- 8784\AA\ during the second. A total of 81 and
45 spectra were obtained during the first and second nights
respectively.

For the blue arm, the standard readout mode was used on the EEV12 CCD
(4k$\times$2k) windowed to 1100\,$\times$\,4200 pixels.  Exposure
times were 180\,s the first night and 300\,s the second with 47\,s
dead time. The spectral resolution corresponds to $38\,\kmsec$ at
$\lambda$4658\AA. The 1200 lines\,mm$^{-1}$ grating R1200B was used to
cover the wavelength range $\lambda\lambda$4253 -- 5058\AA\ enabling
simultaneous monitoring of several He\,{\sc i} lines and \heii. A total
of 132 spectra were obtained during both nights.

After debiasing and flat-fielding the frames by tungsten lamp
exposures, spectral extraction proceeded according to the optimal
algorithm of Marsh (1989).  The data were wavelength calibrated using
CuAr and CuNe arc lamps. The arcs were extracted using the profile
associated with their corresponding target to avoid systematic errors
caused by the spectra being tilted. The spectra were then corrected
for instrumental response and extinction using the flux standard
BD332642 (Stone 1977). The telluric standard HR5718 (Osawa 1959) was
used to correct for the atmospheric water absorption bands.
Uncertainties on every point were propagated through every stage of
the data reduction.

\begin{table}
\centering
\begin{minipage}{84mm}
\caption{Journal of optical observations. $E$ is the cycle number plus binary 
  phase with respect to the ephemeris given by Marsh (1999).}
\label{obs:jo}
\begin{center}
\begin{tabular}{ccccccr}
\hline
\hline
\multicolumn{1}{c}{Date} & \multicolumn{1}{c}{Start} & 
\multicolumn{1}{c}{End} & \multicolumn{1}{c}{Start} & 
\multicolumn{1}{c}{End} & \multicolumn{1}{c}{No. of} \\
\multicolumn{1}{c}{} & \multicolumn{2}{c}{(UT)} & 
\multicolumn{2}{c}{($E -$ 123\,000)} & \multicolumn{1}{c}{frames} \\ \hline
WHT/Red & & & & & \\
26/3/99 & 23.12 & 5.55 & 374.814 & 383.102 & 81 \\
27/3/99 & 22.86 & 5.82 & 405.403 & 414.370 & 45 \\
WHT/Blue & & & & & \\
26/3/99 & 23.12 & 5.57 & 374.815 & 383.121 & 81 \\
27/3/99 & 22.86 & 5.83 & 405.402 & 414.379 & 51 \\
JKT & & & & & \\
26/3/99 & 23.72 & 5.62 & 375.577 & 383.185 & 232 \\
27/3/99 & 22.17 & 4.88 & 404.509 & 413.155 & 224 \\
HST     &       &      & \multicolumn{2}{c}{($E -$ 76\,000)} & \\
4/2/95  & 17.96 & 20.99 & 643.752 & 647.654 & 13 \\ 
\hline

\end{tabular}
\end{center}
\end{minipage}
\end{table}

Simultaneous B-band photometry of \gpc\ was taken during both nights
using a Harris B filter with a Tektronix CCD (1k$\times$1k) on the
1.0\,m Jacobus Kapteyn Telescope (JKT). Exposure times during the
first night ranged between 10 and 40 seconds depending on the seeing.
The seeing worsened during the second night thus the exposure times
had to be increased to values between 50 and 75 seconds.  These data
were used to calibrate the blue spectra in flux. Therefore any
variations present in the blue spectra after this calibrations would
be intrinsic and not due to slit losses or atmospheric variations. We
obtained a total of 118 calibrated spectra after these corrections.

We also present 13 UV spectra of \gpc\ taken with HST on the 4th of
February 1995 using the G140L grating on the High Resolution
Spectrograph (HRS). The wavelength range covered by the spectra is
$\lambda\lambda$1150\AA -- 1440\AA\ with a dispersion of 0.572\,\AA\ 
per diode. Exposure times were of the order of 400\,s. The data were
reduced using the pipeline software. Details of the observing times
and orbits covered are given in Table~\ref{obs:jo}.

\section{Results}
\subsection{Average spectra}

\begin{figure}
\begin{picture}(100,0)(-270,239)
\put(0,0){\includegraphics{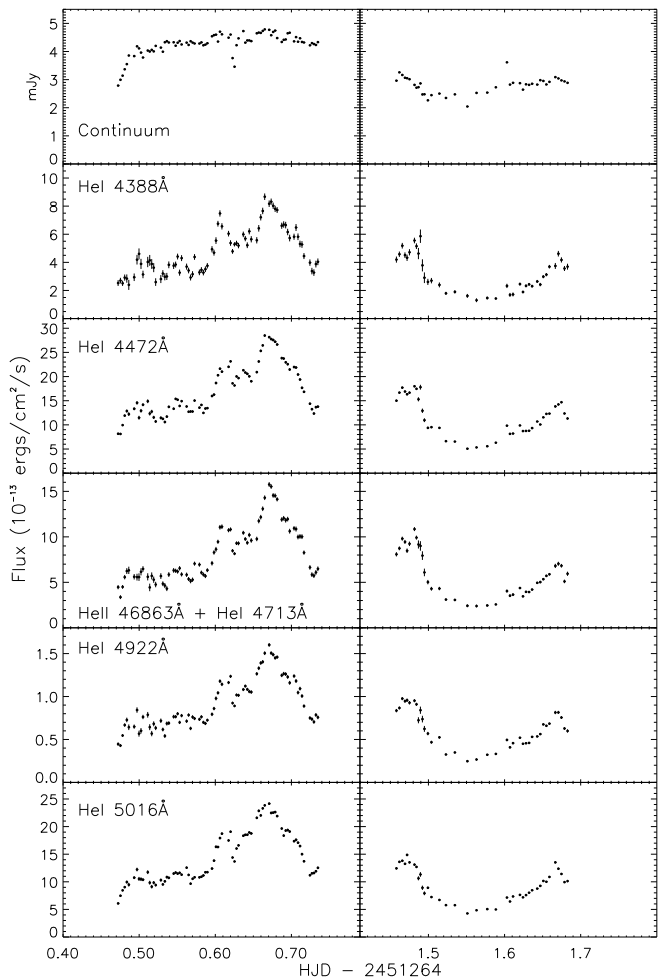}}
\noindent
\end{picture}
\vspace{132mm}
\caption{Light curves of the blue continuum and several emission lines
  found in the blue spectra. GP~Com shows strong variability in all
  the lines. This variability does not seem to be due to its orbital
  motion.}
\label{res:lc}
\end{figure}

The profiles in \gpc\ consist of a classic double-peaked line from the
disc plus a narrow ``spike'' at the centre of each line (see
Fig.~\ref{res:avspall}).  The double-peaked lines are formed by
Doppler shifting with the line wings coming from close to the white
dwarf while the peaks themselves come from the outer edge of the
emission region. Fig.~\ref{res:avspall} presents the average spectra
obtained for each wavelength range observed during this campaign. The
top spectrum can be directly compared with Fig.~1 of Marsh (1999) and
Fig.~1 of Nather et al. (1981). Although Nather et al.'s spectra were
not flux calibrated we can still compare the relative strength of the
central spike with respect to the wings of the line between the three
different sets of data. In Marsh's and Nather et al.'s spectra the
strength of the central spike relative to the wings of the lines is
comparable. In our spectrum the wings of the lines are as strong as
those presented by previous authors but the central spike is $\sim$1.7
times stronger indicating that \gpc\ shows long term variability and
evidence that the spike participates in the variability. In Marsh
(1999) the signal to noise ratio of the data was not high enough to
confirm that the central spike was varying. The central spike is
clearly red-shifted in all emission lines apart from \hee, the
red-shift being different for each line (see Table~\ref{res:cs}) as
already noticed by Marsh (1999).

A new feature present in our data is the absorption located to the
blue of \heii. The fact that this absorption was not present in
Marsh's (1999) spectra indicates that it is a transient feature.

The resolution of the spectrum in the middle panel (wavelength range
$\lambda\lambda$6600 -- 7408\AA) allows us to see the central spike
(red-shifted also in these cases) for three more He\,{\sc i} lines.
The spectrum in the lower panel shows a complicated combination of
many N\,{\sc i} lines, but none of any other element.

\subsection{Flares}
\label{op:flares}

\begin{figure}
\begin{picture}(100,0)(-270,239)
\put(0,0){\includegraphics{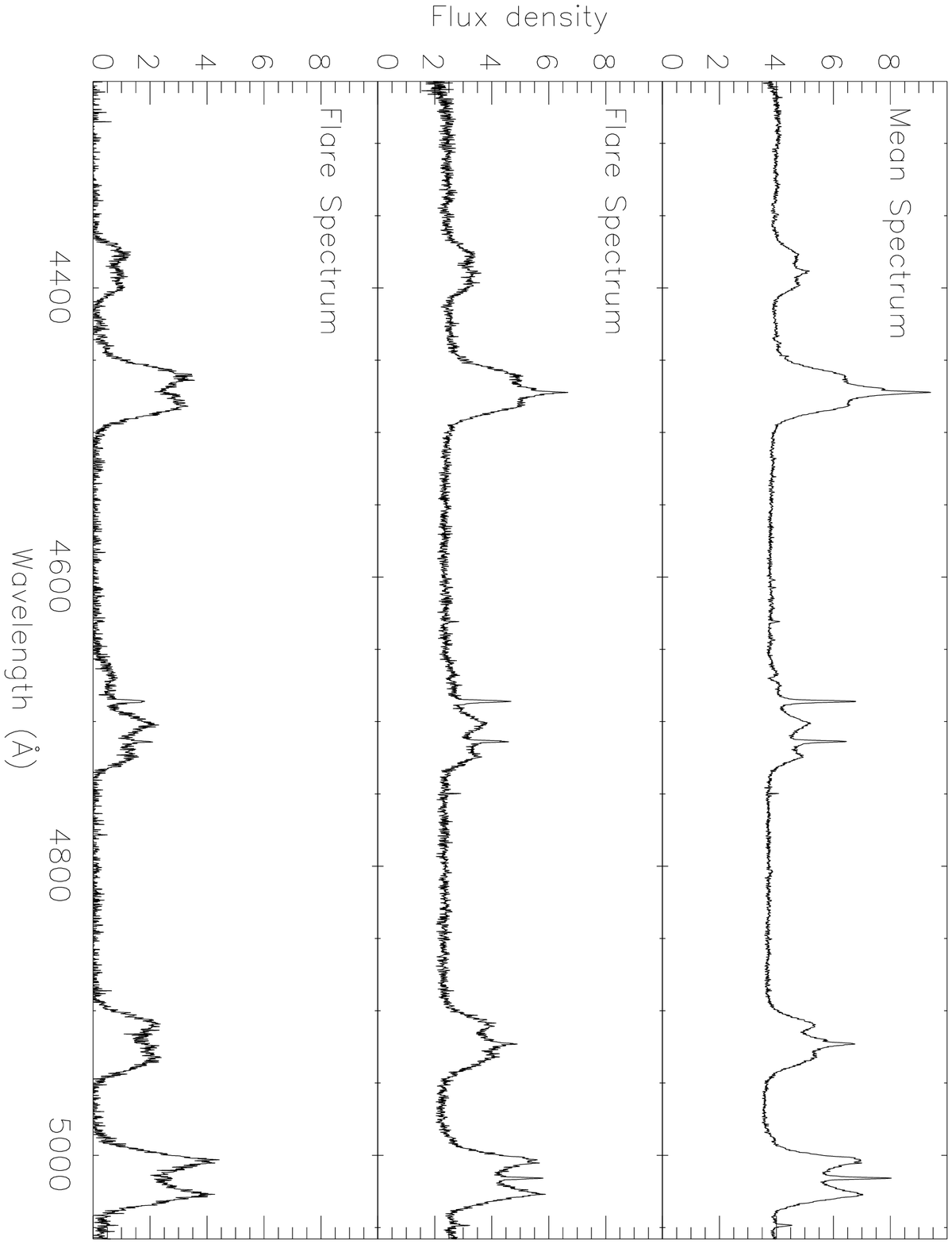}}
\put(0,0){\includegraphics{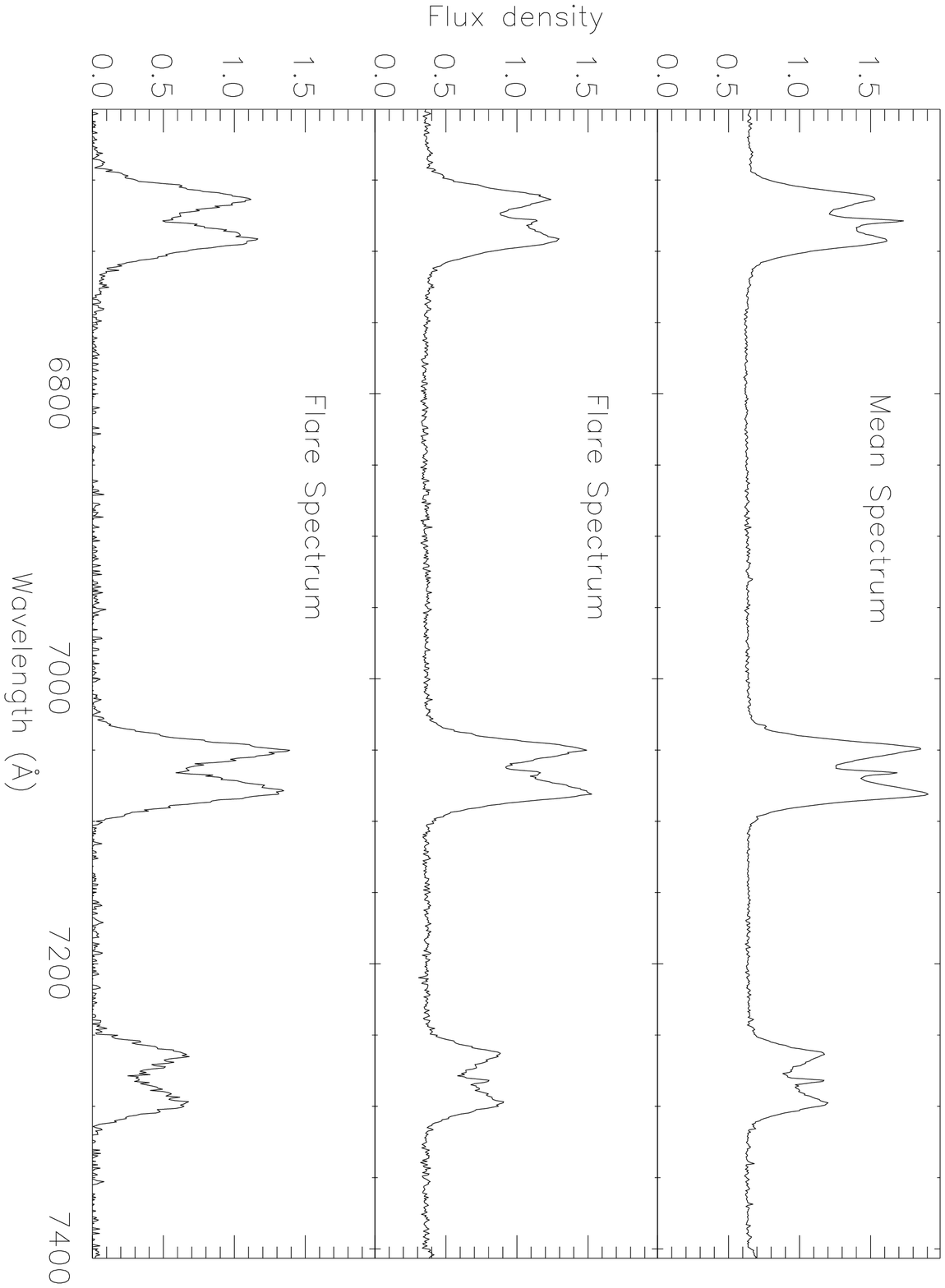}}
\put(0,0){\includegraphics{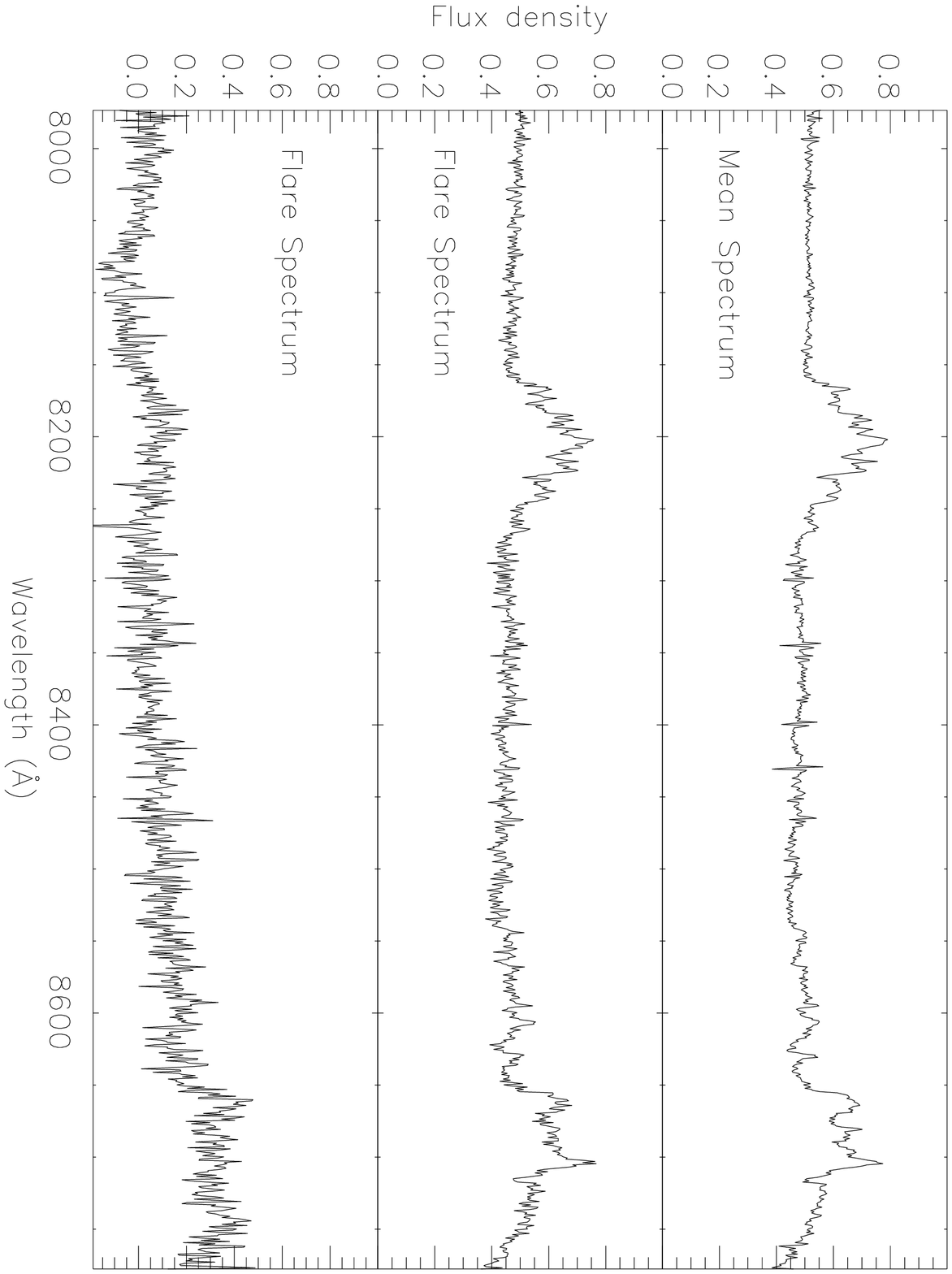}}
\noindent
\end{picture}
\vspace{200mm}
\caption{The panels show the mean \gpc\ spectrum (top), the
  flare spectrum obtained from the raw spectra (middle) and the flare
  spectrum obtained from the continuum normalised spectra (bottom).}
\label{res:flb}
\end{figure}

\gpc\ displays strong flaring in X-ray (van Teeseling \&\ Verbunt
1994), UV (Marsh et al. 1995) and optical (Marsh et al. 1991)
wavelengths.  This variability is clearly present in our data, mainly
in the emission lines. Fig.~\ref{res:lc} shows the light curves of the
continuum and blue emission lines over this two day run. The blue data
($\lambda\lambda$4253 -- 5058\AA) presented in this work are
photometric, as we used simultaneous photometry to correct from
atmospheric variations in the spectra, therefore any variation
observed in the data is intrinsic. This is confirmed by the fact that
we see the same variations in all the blue lines (Fig.~\ref{res:lc}).

The lightcurves obtained for both nights are very different. The
variability seen in the lines and the continuum is also different, the
continuum varying less significantly than the lines. For the emission
lines, we see that during the first night \gpc\ shows short term
flares that last of the order of 15 to 70\,min. These flares are
present throughout the entire night. During the second half of the
first night these short timescale flares are superimposed to a longer
timescale brightening in the lines that lasts of the order of 3 hours.
On the other hand, during the second night, although we also see short
timescale flares, these take place mainly at the beginning and at the
end of the observations with very little variability in between. We
searched for periodicities in the lightcurves of the emission lines by
using a floating mean periodogram (Cumming et al. 1999) and found
nothing significant. All the emission lines show similar variability.
  
As for the continuum, the variability observed is smaller than that
seen in the emission lines, 10 to 20 per cent during the first night
and up to 30 per cent during the second night compared with variations
of up to 80 per cent in the emission lines. During the second night
the variations seen in the continuum seem to follow those seen in the
lines, i.e. a flare at the beginning of the observations followed by a
decrease and perhaps a general increase at the end of the night. This
is not so clear during the first night, although the initial increase
seen in the continuum and the decrease at HJD - 2451264 = 0.63 match
the behaviour seen in \hee. A search for periodicities in the
continuum lightcurves, using the same method as for the emission lines
yields no significant period, in particular the orbital period is not
present.
  
We used the same method described in Marsh et al. (1995) to obtain the
spectrum for the flaring component. This consists of modelling the
spectra as the sum of two components, one of them being the mean
spectrum and the other a multiple of the flare component. The model
fitting is carried out iteratively until the two spectra and the
multipliers are optimised by minimising $\chi^2$. All the spectra are
used to obtain the mean and the flare components because during the
calculations each spectrum is weighted according to its total flux
giving the highest signal to noise ratio in the flare spectrum.  The
average and flare components of the spectra are shown in
Fig.~\ref{res:flb}. The calculation of the flare spectrum was made by
using the raw spectra (middle panel) and the continuum normalised
spectra (bottom panel). By using the continuum normalised spectra we
can investigate whether the broad component of the lines and the
central spike vary in the same way as the continuum.

\begin{figure}
\begin{picture}(100,0)(-270,189)
\put(0,0){\includegraphics{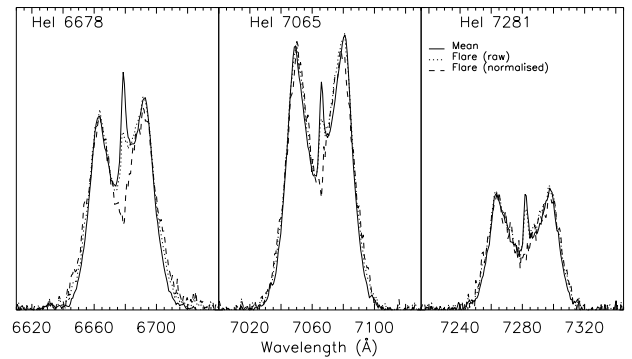}}
\noindent
\end{picture}
\vspace{48mm}
\caption{Average and flare components for the lines \hef, \heg\ and
  \heh. The flare components (dotted and dashed lines) have been
  scaled to match the average component of the lines. Notice that the
  flare profiles are broader than the average profiles indicating that
  they form at higher velocity regions.}
\label{res:flc}
\end{figure}

The flare spectra obtained in these two cases are noticeably
different, i.e. the central spike does not seem to be present in many
of the emission lines in the flare spectrum obtained from the
continuum normalised spectra but it is present in all lines for the
raw spectra. The broad component of the helium lines is comparable in
both cases which indicates that they do not vary in a similar way to
the continuum, i.e. broad line and continuum variations seem to be
independent. On the other hand the central spike does look different
in both flare spectra. For \heii\ and \hec\ the central spike is
$\sim$1.5 times stronger in the raw spectrum, and for all the other
helium lines the central spike is present. This indicates that the
central spike component varies in a similar way, but not identical, to
the continuum. Where this variability is produced is still an unknown
but we can at least say that the variability in the broad lines has a
different origin to that in the continuum and the central spike.  If
we assume that the optical variability is the result of irradiation by
the emitted X-rays, the flare spectra suggest that the broad component
of the lines is formed in an optically thin region, more responsive to
irradiation, whereas the continuum and the central spike have their
origin in a region less responsive to irradiation, probably optically
thick.

The behaviour seen in the N\,{\sc i} lines is different to that of the
helium lines. In this case the lines also contribute to the flare
spectrum but their variability is not independent from the continuum
variability. Continuum and lines vary in similar ways which is why we
see very little signal in the flare spectrum obtained from the
continuum normalised spectra.

The main question we wanted to answer in this section was whether the
central spike contributed to the flare spectrum as suggested by Marsh
(1999). From Fig.~\ref{res:flb} we see that this is the case.  From
this we conclude that the origin of the central spike seems to be the
accreting star and not a nebula around the system.

Another important difference found between the average and flare
spectra can be seen in Fig.~\ref{res:flc} where we have plotted both
components for \hef, \heg\ and \heh. In all cases the flare component
is broader than the average component indicating that flaring occurs
mainly in high velocity regions, i.e. the inner regions of the disc.
Table~\ref{res:fwzi} gives a list of the change in full width at zero
intensity (FWZI) between the mean and the flare spectrum obtained from
the continuum normalised data. FWZI changes between the mean spectra
and the flare spectra obtained from the raw data are about half those
values given in Table~\ref{res:fwzi}.

\begin{table}
\centering
\begin{minipage}{84mm}
\caption{Changes in FWZI between the helium lines in the mean and
  flare spectra. The flare spectrum used for these measurements was
  that obtained from the continuum normalised spectra. Changes in FWZI
  between the mean and flare spectra obtained from the raw data are
  about 0.5 times those shown.}
\label{res:fwzi}
\begin{center}
\begin{tabular}{lcc}
\hline
\hline
Line  & \multicolumn{2}{c}{$\Delta \rm FWZI$}\\
 & km s$^{-1}$ & \AA \\
\hline
He\,{\sc i}\,4388  & 399 & 5.8\\
He\,{\sc i}\,4472  & 526 & 7.8\\
He\,{\sc ii}\,4686 + He\,{\sc i}\,4713 & 353 & 5.5\\
He\,{\sc i}\,4922 & 149 & 2.4\\
He\,{\sc i}\,5016 & 318 & 5.3\\
He\,{\sc i}\,6678 & 455 & 10\\
He\,{\sc i}\,7065 & 156 & 3.7\\
He\,{\sc i}\,7281 & 84  & 2\\
\hline
\end{tabular}
\end{center}
\end{minipage}
\end{table}

\subsection{Central spike}

Once we have established that the central spike contributes to the
flaring component of the spectra and therefore that its origin is
probably the white dwarf, it is of great interest to measure its
radial velocity amplitude to further constrain its origin. We carry
out this measurement in the same way described by Marsh (1999), by
using multi-Gaussian fitting to the profile of the line. We used the
raw spectra to carry out the fitting, therefore the mean and flare
components are fitted simultaneously. Each emission line is fitted
with four Gaussians, one for each wing, one for the bright spot
component and one for the central spike. In the case of \heii\ only
one Gaussian was used to fit the profile as the right wing was heavily
contaminated by \hec, the left wing showed strong absorption and the
bright spot emission was very faint (see Fig.~\ref{res:bluedopp}). For
\heb\ and \hed\ two Gaussians instead of one were used to fit the
central spike component as this included the satellite feature
discussed in the following section. This was not possible for \hea\ as
the line is significantly fainter than \heb\ and \hed. The values of
the fitting parameters that most interest us are those measured for
the central spike component.  By fitting all the spectra
simultaneously we obtained an optimal value for the width of the spike
component.  We then maintained this parameter fixed and left two
parameters variable, the height of the Gaussian and its offset from
the rest wavelength. The parameters for the other Gaussians used were
left free. The velocities measured for the central spike (equivalent
to the offsets) were then fitted with a sinusoidal function of period
equal to \gpc's orbital period, $\rm V = \gamma + \rm K_{\rm
  cs}\sin(2\pi(\phi - \phi_0))$, where $\gamma$ is the systemic
velocity, $\rm K_{\rm cs}$ is the radial velocity semiamplitude of the
central spike component, $\phi$ is the orbital phase and $\phi_0$ is
the phase at which the radial velocity semiamplitude is zero. This is
equivalent to fitting the data with a function $\rm V = \gamma - \rm
V_{\rm X}\cos{2\pi\phi}+\rm V_{\rm Y}\sin{2\pi\phi}$ where $\rm V_{\rm
  X}$ and $\rm V_{\rm Y}$ are just a combination of the radial
velocity semiamplitude and the orbital phase. We present the fitting
parameters in Table~\ref{res:cs}. The problem of the redshift of \hee\ 
being significantly smaller than that of the other lines found by
Marsh (1999) remains in our data.

\begin{table}
\centering
\begin{minipage}{84mm}
\caption{Velocity parameters of the central spike for some of the
  emission lines.}
\label{res:cs}
\begin{center}
\begin{tabular}{lccc}
\hline
\hline
Line & $\gamma$ & $\rm V_{\rm X}$ & $\rm V_{\rm Y}$\\
     & km\,s$^{-1}$ & km\,s$^{-1}$ &km\,s$^{-1}$\\
\hline
He\,{\sc i}\,4388 & -- & -- & --\\
He\,{\sc i}\,4472 & 47.00 $\pm$ 0.80 & 13.43 $\pm$ 1.12 & 5.37 $\pm$ 0.58\\
He\,{\sc ii}\,4686 & 20.67 $\pm$ 0.29 & 10.32 $\pm$ 0.42 & 0.75 $\pm$ 0.09\\
He\,{\sc i}\,4713 & 38.34 $\pm$ 0.54 & 11.82 $\pm$ 0.76 & 3.43 $\pm$ 0.33\\
He\,{\sc i}\,4922 & 64.91 $\pm$ 0.89 & 10.71 $\pm$ 1.41 & 0.17 $\pm$ 0.22\\
He\,{\sc i}\,5016 & 8.85 $\pm$ 0.53 & 10.36 $\pm$ 0.73 & 2.68 $\pm$ 0.30\\
He\,{\sc i}\,6678 & 29.04 $\pm$ 0.68 & 12.37 $\pm$ 1.03 & 0.38 $\pm$ 0.18\\
He\,{\sc i}\,7065 & 36.62 $\pm$ 0.61 & 11.06 $\pm$ 0.88 & 0.26 $\pm$ 0.16\\
He\,{\sc i}\,7281 & 37.17 $\pm$ 1.00 & 13.37 $\pm$ 1.42 & 2.24 $\pm$ 0.46\\
\hline
\end{tabular}
\end{center}
\end{minipage}
\end{table}

\subsubsection{The origin of the double-peaked central spike}

The lines \hea, \heb\ and \hed\ share a feature not seen in other
lines: the central spike has a satellite blueward of the main peak.
In \hea, this satellite is comparable to the main peak whereas for the
other two lines the satellite is weaker than the main peak.  These can
be seen most clearly in the trailed spectra of
Fig.~\ref{res:bluedopp}.  We believe these features are the result of
Stark effect.

The presence of charged particles in the gas makes Stark broadening an
important effect to look for in the spectra of \gpc. In the case of
hydrogen, emission and absorption lines split symmetrically and, if
the electric field is not very strong, the hydrogen line will appear
broadened. In the case of neutral helium, the presence of the charged
particles in the plasma causes the upper and lower energy states of a
permitted transition to mix with their neighbouring states. This
results in forbidden neutral helium transitions being excited
(Beauchamp \& Wesemael 1998). A given neutral helium forbidden
transition is therefore associated with a permitted one. These
forbidden neutral helium lines have been observed in many helium rich
white dwarfs (DB white dwarfs; Beauchamp et al. 1995, 1997).

\begin{table*}
  \caption{Columns 1 and 2 show the wavelengths of the forbidden and
    permitted helium transitions respectively. The expected
    separations in wavelength between forbidden and permitted
    transitions are shown in column 3. Column 4 gives a measurement of the
    expected relative strength of each forbidden transition. These
    first four columns are taken from Table 1 of Beauchamp \& Wesemael
    (1998). The last column presents the separations we measure
    between peaks in our data. T$_4$ is the local temperature in units
    of 10,000\,K.} 
\label{res:BW}
\begin{center}
\begin{tabular}{lllll}
\hline\hline
Forbidden& Permitted& $\lambda_{\rm f} - \lambda_{\rm p}$ (\AA)& $\log(\chi_{\rm forb}/\kappa_{\rm \nu})$ &
$\Delta \lambda$\\
\hline
4387 & 4388 & $-$0.54 & 5.67$-$0.13/T$_4$ & 1.25$\pm$0.28\\
4470 & 4472 & $-$1.48 & 5.18 & 2.57$\pm$0.28\\
4921 & 4922 & $-$1.32 & 4.83$-$0.13/T$_4$ & 2.86$\pm$0.27\\
4383 & 4388 & $-$4.65 & 3.74$-$0.13/T$_4$ & --\\
4911 & 4922 & $-$11.2 & 3.04$-$0.13/T$_4$ & --\\
4517 & 4472 & +46.0 & 2.26 & --\\
5042 & 5016 & +26.4 & 1.44+0.17/T$_4$ & --\\ 
\hline
\end{tabular}
\end{center}
\end{table*}

Beauchamp \& Wesemael (1998) in their Table 1 give a list of forbidden
transitions that might be found in helium rich stars and a measure of
the probability with which we should see them in their spectra. The
order in which the forbidden components appear in the list is the
order in which they should become observable. We present an extract of
their results together with some measurements from our data in
Table~\ref{res:BW}. The table only presents the predictions for the
lines covered by our wavelength range. The first and second columns
give the values of the forbidden and permitted transitions
respectively. The third column shows the separation in \AA\ expected
between both components. The fourth column shows the ratio of the line
opacity associated with the forbidden component, $\chi_{\rm forb}$, to
the local continuum opacity, $\kappa_{\rm \nu}$, as a measure of the
relative strength of the forbidden component in the spectrum. If one
of the forbidden transitions listed in the table is present in the
spectra then all forbidden transitions with values of $\chi_{\rm
  forb}/\kappa_{\rm \nu}$ larger than that of the observed transition
should also be present.  The last column in Table~\ref{res:BW} gives
the separation between the two components of the central spike seen
our spectra.  According to Beauchamp \& Wesemael's (1998) argument,
the fact that we see two components in \hed\ implies that we should
also see them in \heb\ and \hea\ which agrees with our observations.
The predicted separations between the forbidden and permitted
transitions do not agree but they seem to follow the same trend, i.e.
the separation between both components in \hea\ is significantly
smaller than for the other two lines, and for these the separations
are comparable. In fact, the predicted and observed wavelength
separations seem to differ by a factor of $\sim$2. We do not have an
explanation for this discrepancy.

The relevance of this result is enormous because it could cause the
lines to be shifted around by different amounts, and also, as there is
an asymmetry involved, it could be the explanation for the different
redshifts observed for the central spike in different lines.

A direct consequence of this claim is the prediction of the presence
in the spectrum of \gpc\ of the forbidden helium components that have
larger values of $\chi_{\rm forb}/\kappa_{\rm \nu}$ than the ones
observed in our data. These lines, given in Beauchamp \& Wesemael's
(1998) table, are 3705, 4009, 3820, 4144 and 4026 \AA. It is worth
pointing out that the separations expected between the forbidden and
permitted components for He\,{\sc i}\,3705, 4009, 3820 and 4144 \AA\ 
are smaller than those measured here and higher resolution spectra
will probably be required to be able to resolve both components. In
the case of the 4026 \AA\ line, the separation expected is larger than
than of \hea\ (which is in the limit of our resolution) so spectra of
similar resolution to the ones presented here will suffice to test our
prediction.

According to Beauchamp \& Wesemael (1998), the profiles of these
neutral helium forbidden components are good indicators of the density
and gravity of the medium in which they form. The denser the medium
the more symmetrical and wider the profiles. Also the higher the
gravity the more symmetric the profiles appear. It would be very
interesting to apply model atmosphere codes that have been developed
for DB white dwarfs to \gpc\ and similar systems to calculate the
density and gravity of the region where the central spike forms.

This study should be carried out also for the recently discovered twin
system of \gpc, CE315. Being so similar to \gpc\ we might expect to
find that the Stark effect is also of importance and that the neutral
helium forbidden components are present in its spectrum.

\subsection{Doppler maps}
\label{dop}

\begin{figure*}
\begin{picture}(100,0)(-270,250)
\put(0,0){\includegraphics{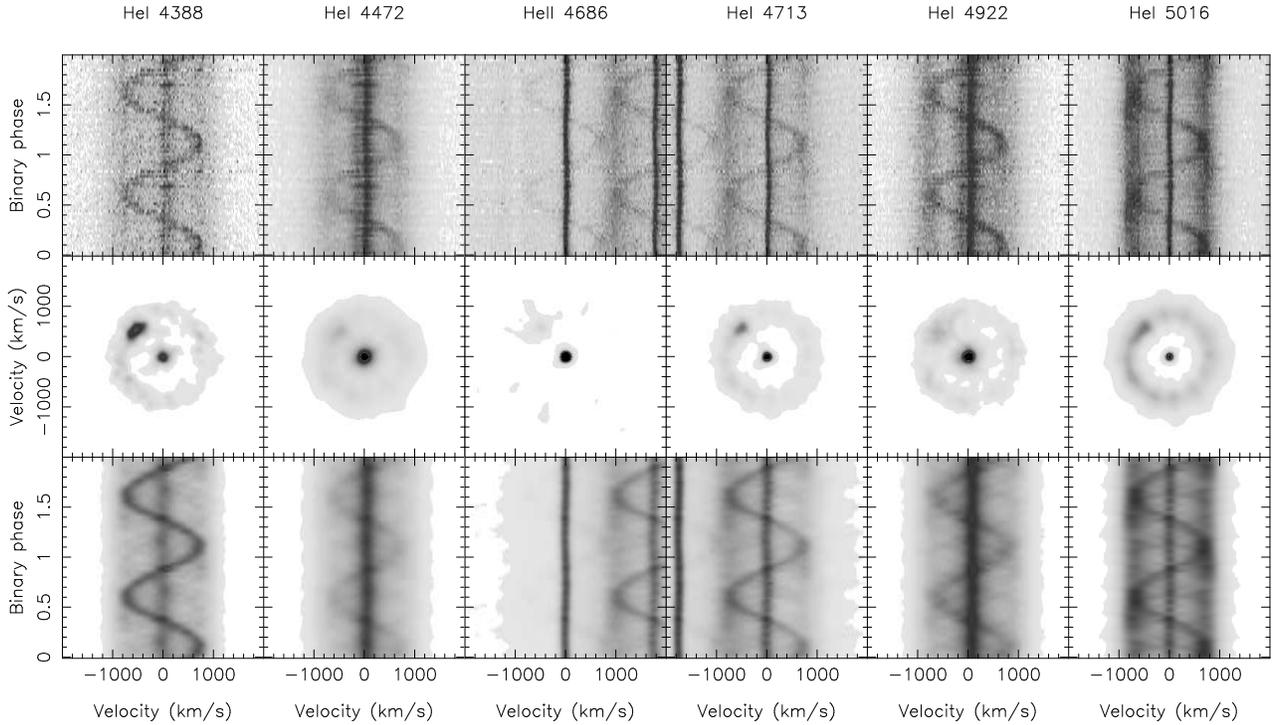}}
\noindent
\end{picture}
\vspace{94mm}
\caption{Trails of the lines (top
  panels), Doppler maps (middle panels) and trails of the lines
  computed back from the maps (bottom panels). We have corrected the
  ephemeris given by Marsh (1999) so the modulation of the central
  spike corresponds to the motion of the compact object. Emission is
  clearly seen in the position corresponding to the bright spot.}
\label{res:bluedopp}
\end{figure*}

\begin{figure*}
\begin{picture}(100,0)(-270,250)
\put(0,0){\includegraphics{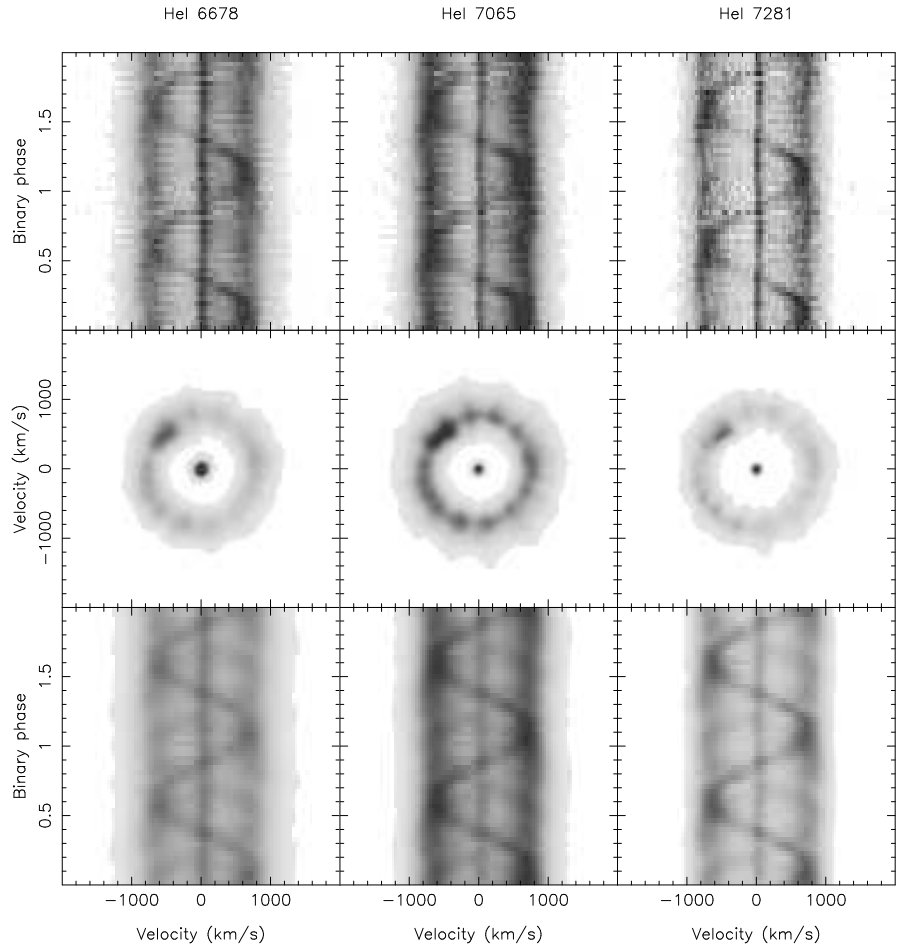}}
\noindent
\end{picture}
\vspace{121mm}
\caption{Same as for Fig.~\ref{res:bluedopp} but for the three He\,{\sc i}
  lines observed in the red. Notice that the intensity variations
  seen along the accretion disc on the three lines and more
  significantly in \heg\ are the result of poor orbital phase
  coverage.}
\label{res:reddopp}
\end{figure*}

We present the trails of the helium emission lines present in the
spectra in the top panels of Figs.~\ref{res:bluedopp} and
\ref{res:reddopp}. The spectra have been rescaled to take away the
line flux variations resulting from GP Com's flaring. The ephemeris
calculated by Marsh (1999) is not precise enough to be extrapolated
thus we used a corrected ephemeris to bin and fold the data:
\begin{equation}
\rm HJD = 2451200.2745 + 0.0323386 \rm E.
\end{equation}
This corrected ephemeris is Marsh's (1999) shifted by 0.25 of an orbit
so the modulation of the central spike corresponded to the motion of
the compact object. We present two orbital cycles for each line. In
all lines we see the sinusoidal component that suggested the binary
nature of \gpc. It zig zags between the two wings of the lines. The
central spike appears in the trails as the vertical line fixed at low
velocities. Doppler maps of the lines obtained using the maximum
entropy method (MEM) are presented in the middle panels of
Figs.~\ref{res:bluedopp} and \ref{res:reddopp}.  The central spike
maps into the low velocity emission in the centre of the maps.  The
double wings located at $\sim$ 800 \kmsec\ either side of the rest
wavelength map into an accretion disc around the central emission.
Lastly, the ``sinusoidal'' component maps into emission coming from a
region located on the accretion disc. This emission had previously
been interpreted as coming from the bright spot (Marsh 1999). We
notice that the bright spot shows a complex structure for some lines
(i.e. \hef, \heg, \heh), stretched along the accretion disc in the
maps. This behaviour, already seen by Marsh (1999), indicates that the
bright spot is actually moving in a semi-sinusoidal fashion probably
due to phase-dependent visibility in the emission regions, i.e at some
phases we see emission from the stream and at others emission from the
disc near the stream. In the next section we present a variation of
Doppler tomography that takes into account flux variability in the
lines. By means of this new technique we see that the behaviour of the
bright spot is rather unusual.

\subsubsection{Eccentricity of the accretion disc}

Another interesting effect seen in some of the lines (\hef, \heg,
\heh\ mainly) is that the accretion disc does not appear circular in
the maps but somewhat elliptical. Once the motion of the white dwarf
has been subtracted, a circular accretion disc would be seen in the
trails of the lines as two vertical lines situated at the same
distance at both sides of the rest wavelength. If the disc were
elliptical, the vertical lines would not be at the same distance from
the rest wavelength at all times. They would show higher radial
velocities at orbital phases that correspond to a higher distance from
the centre of the map. If we subtract the average spectrum from all
spectra, we would be left with only the deviations from the circular
disc.

\begin{figure}
\begin{picture}(100,0)(-270,250)
\put(0,0){\includegraphics{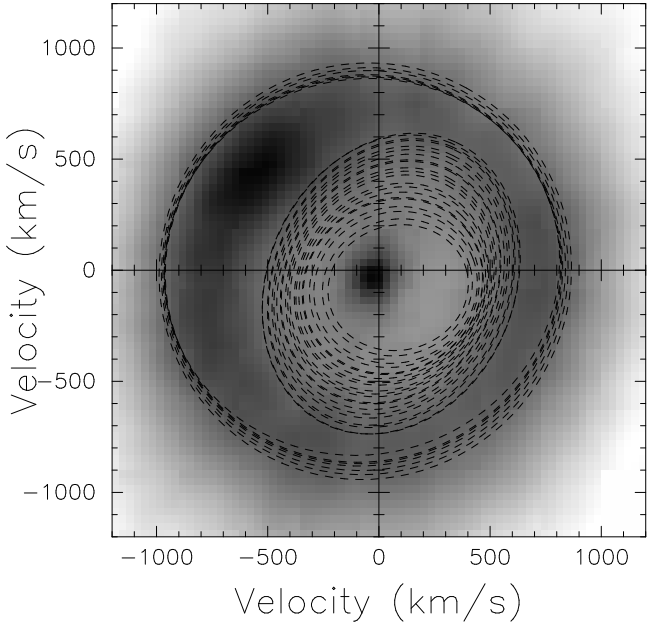}}
\noindent
\end{picture}
\vspace{85mm}
\caption{Doppler tomogram obtained by adding up the signal from the
  three emission lines in the red spectrum: \hef, \heg, and \heh. Fits
  with ellipses to regions of equal flux were performed and plotted in
  the tomogram. }
\label{res:elli1}
\end{figure}

\begin{figure}
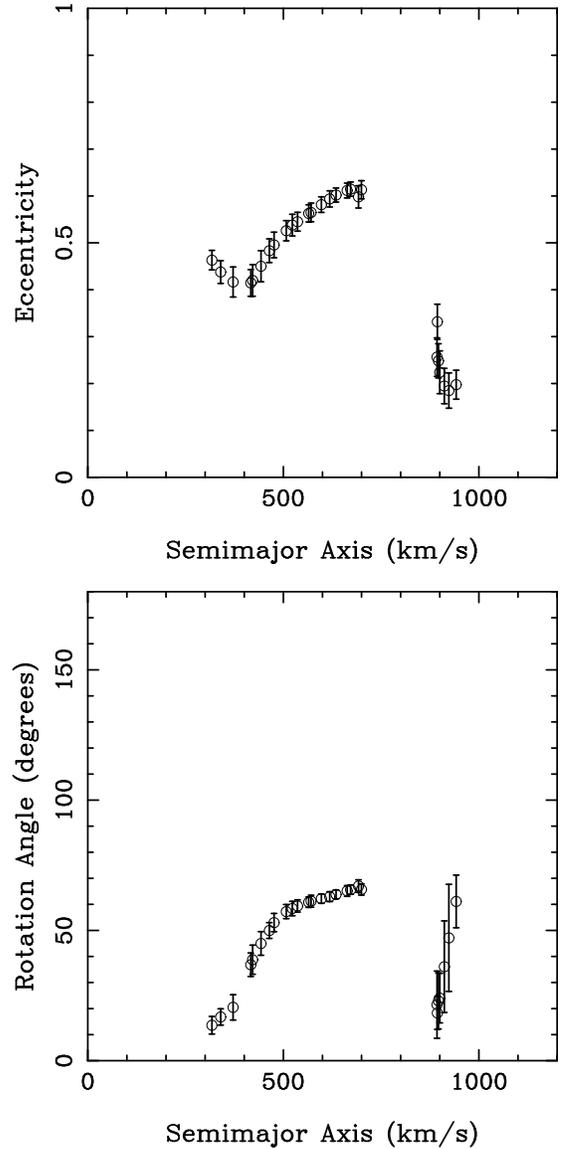

\begin{picture}(100,0)(-270,250)
\put(0,0){\includegraphics{h4241f10.ps}}
\put(0,0){\includegraphics{h4241f11.ps}}
\noindent
\end{picture}
\vspace{155mm}
\caption{The top panel represents the eccentricity of the ellipses
  fitted to the Doppler map in Fig.~\ref{res:elli1}. As the ellipses
  get closer to the outer rim of the disc the eccentricity becomes
  constant with a value of $\sim$0.6. The ellipses fitted to the inner
  regions of the disc start with a value for the eccentricity of 0.3
  and then become more circular as they get closer to the white dwarf.
  The bottom panel represents the angle of the semimajor axis in the
  ellipse versus the length of the semimajor axis. The angle is
  measured from the positive x velocity axis and increases
  anticlockwise. The angle increases as the ellipses get closer to the
  outer rim of the disc.}
\label{res:elli2}
\end{figure}

To confirm the disc's ellipticity we measured the eccentricity in
different regions of the Doppler map. In Fig.~\ref{res:elli1} we
present a Doppler tomogram for \gpc\ obtained after adding up the
three emission lines in the red part of the spectrum, \hef, \heg\ and
\heh. Before we added the spectra for the three lines, the motion of
the white dwarf was subtracted in each case using the solutions given
in Table~\ref{res:cs}.
We then fitted elliptical curves to the equal flux contours in two
different zones of the map. The procedure is described in Unda-Sanzana
et al. (in preparation) but is briefly outlined here. It consists of
two stages.  After masking undesired regions (e.g. bright spot) the
first stage is the selection of points at a certain level of flux,
automatically picking them in such a way that the selection lies in a
closed and continuous contour surrounding a point specified by the
user (e.g. the origin). This is very straightforward for most
``smooth'' cases, but it is problematic for cases when large flux
variations occur from pixel to pixel, typically near the tomogram's
cusps or well into its outer regions. In those cases the selection
would have to be done by hand by the user, so it is most reliable to
exclude those regions from the process. The second stage is the use of
the analytical ellipse-specific fitting algorithm by Fitzgibbon et al.
(1999), in order to fit an ellipse to the selection. The outcome of
the fitting is a number of parameters characterising the ellipse,
which are later translated into physical quantities such as the
rotation angle, size of semiaxes, etc.  The selection is then
bootstrapped and the fitting process repeated a large number of times
in order to obtain uncertainties for the parameters. The first zone to
which ellipses were fitted is the low velocity region in the map which
corresponds to the outer regions of the accretion disc. The second
zone is the high velocity region in the map which corresponds to the
inner accretion disc. The top panel of Fig.~\ref{res:elli2} presents
the eccentricity of the ellipses plotted in Fig.~\ref{res:elli1}
versus semimajor axis. The central region of the accretion disc, where
most of the emission is coming from, has not been fitted with ellipses
because, as explained above, the first stage of the process is
unreliable under the large flux variations of this zone. These large
flux variations in the disc can be produced by the presence of the
elongated bright spot or by the phase coverage of the data introducing
artifacts.

The eccentricities of the ellipses fitted to the first zone give a
consistent value of $\sim$0.6 which indicates that the outer regions
of the disc are not circularly symmetrical in flux. The bright spot
was masked when we performed the fits so the asymmetry does not arise
from its presence in the outer rim of the disc. For the second zone,
the eccentricity near the boundary layer is $\sim$0.3 and then it
decreases to 0 soon after. This decrease is probably the result of
getting closer to the white dwarf.

The bottom panel of Fig.~\ref{res:elli2} shows the angle of the
semimajor axis for each ellipse. The angle is measured from the
positive x velocity axis and increases anticlockwise. We see that for
the first zone the rotation angle increases for each ellipse as we get
closer to the outer rim of the disc. As for the second region, the
angle of the ellipses also increases as we get closer to the white
dwarf.

According to Paczynski (1977), the maximum size for a very small
pressure and viscosity accretion disc can be approximated by the
orbits of a test particle in the restricted 3-body problem.  In the
case of \gpc, where $\rm q = \rm M_{2}/\rm M_{1}$ and $\mu = \rm
M_{2}/(\rm M_{1} + \rm M_{2})$ have very close values, i.e. $\sim$0.02
(Marsh 1999) we would expect the outermost stable orbit to be
elliptical with eccentricity e$\sim$0.5 (using the closer value for
$\mu$ in Paczynski 1977). The eccentricity we measure in
Fig.~\ref{res:elli2} is somewhat larger than this, $\sim$0.6. If
3-body effects are the cause of the eccentricity seen in the disc, we
would expect the semimajor axis of the elliptical disc to be
perpendicular to the line that joins both components of the system in
spatial coordinates. In velocity space this translates into the
semimajor axis of the elliptical disc being parallel to the line that
joins both components of the system, the white dwarf and the donor
star. This argument implies that the donor star would be situated
about 70$^{\circ}$ anticlockwise of the positive x axis in the
velocity maps presented in Figs.~\ref{res:bluedopp}, \ref{res:reddopp}
and \ref{res:elli1}, according to the rotation angle measured in the
bottom panel of Fig.~\ref{res:elli2}. Notice that we choose the
closest orbit to the visible disc as the largest stable orbit. We do
not find any significant emission other than that coming from the
accretion disc at such position which could just be the result of the
donor star being very dim. If the donor star were placed in that
position in the Doppler maps, the angle between the positions of the
donor star and the hot spot would be between $\sim$58$^{\circ}$ and
$\sim$79$^{\circ}$. We have calculated the trajectory of the stream in
velocity space for different values of the mass ratio $q$ and find
that for the bright spot to be between 58$^{\circ}$ and 79$^{\circ}$
from the donor star, the size of the accretion disc would be as given
in Table~\ref{res:qbspos}. In the table we also present the maximum
radius of the largest stable orbit according to Paczynski (1977) and
Henon (1969) which corresponds to the semi-major axis in an elliptical
orbit. From the values in Table~\ref{res:qbspos} we deduce that it
could be possible that the eccentricity seen in the accretion disc
would be due to 3-body effects as the values of the radii of the
largest stable orbits and the positions of the bright spot are
comparable for all the mass ratios considered (more so for values
obtained according to Henon 1969). This argument implies that the
ephemeris given in Sect.~\ref{dop} might be incorrect and therefore
the absolute orbital phases used throughout the paper are uncertain.

\begin{table}
\caption{Position of the bright spot and maximum radius of the largest stable
  orbit according to Paczynski (1977) (first value) and Henon (1969)
  for five different mass ratios. The position of the bright spot was
  calculated assuming that the angle between the donor star and the
  spot is between 58$^{\circ}$ and 79$^{\circ}$. Distances are given
  in terms of a, the distance between both stars in the system, and
  are measured from the compact object.}
\label{res:qbspos}
\begin{center}
\begin{tabular}{ccc}
\hline\hline
Mass ratio & bright spot (a) & Max. stable orbit (a)\\
\hline
0.01 & 0.474 - 0.387 & 0.559;0.472\\
0.02 & 0.452 - 0.345 & 0.537;0.466\\
0.03 & 0.443 - 0.325 & 0.521;0.459\\
0.04 & 0.438 - 0.312 & 0.509;0.454\\
0.05 & 0.436 - 0.304 & 0.498;0.449\\
\hline
\end{tabular}
\end{center}
\end{table}

It is worth pointing out that Unda-Sanzana et al. (in preparation)
have carried out a similar study for the CV U Gem and find that
although the accretion disc is eccentric, the semi-major axis of the
ellipses that fit the disc is not aligned with the line that joins the
compact object and the donor star in the velocity maps. In U Gem the
donor star is clearly seen in the spectra and therefore in the Doppler
maps.

\subsubsection{Modulation Doppler maps}

\begin{figure*}
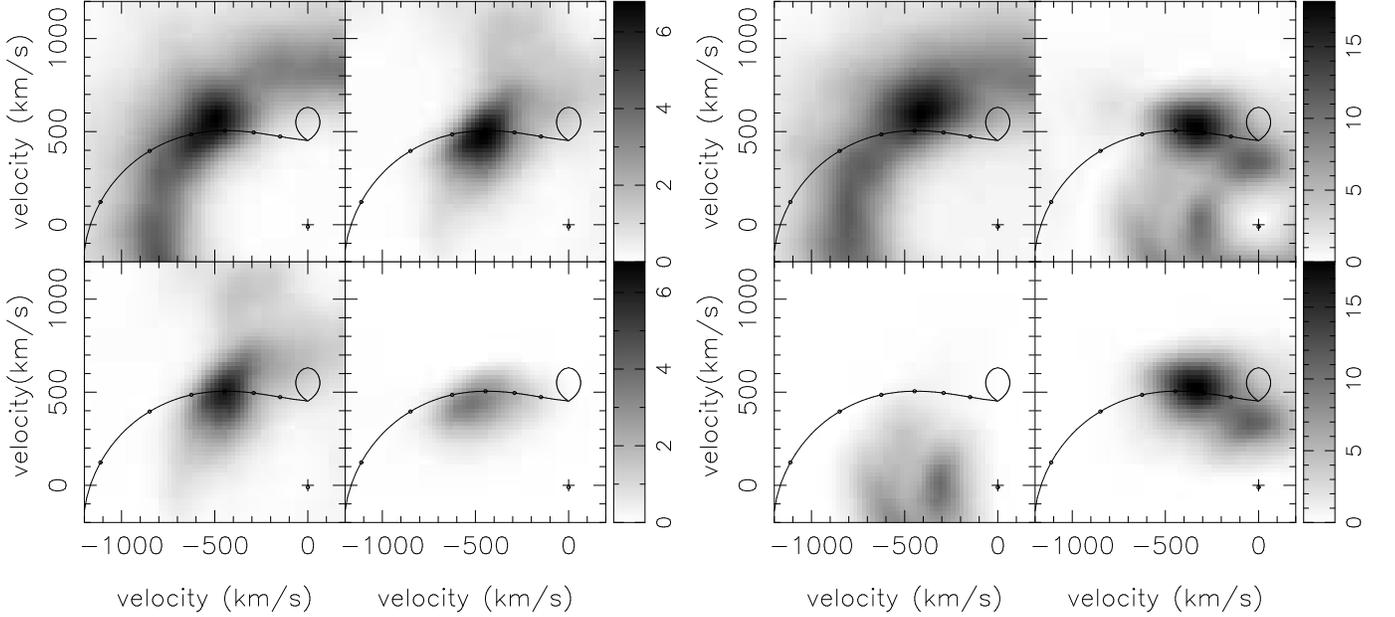

\begin{picture}(0,0)(450,450)
\put(0,0){\includegraphics{h4241f12.ps}}
\put(0,0){\includegraphics{h4241f13.ps}}
\noindent
\end{picture}
\vspace{85mm}
\caption{Modulation Doppler images for \hee\ (left) and \hef\ (right). Only the
  parts of the maps close to the bright spot are displayed. The top
  left panels of both figures represent the constant part of the line
  emission, the top right the total amplitude of the modulated part of
  the emission. The grayscale wedge denotes the amplitude scale in
  percentage. The bottom two panels of both figures split the variable
  part in its cosine (left) and sine contributions. A $q=0.02$
  ballistic stream trajectory is also plotted for reference. }
\label{res:modmaps}
\end{figure*}

We have mentioned in Sect.~\ref{dop} that the bright spot emission
shows deviations from a sinusoidal curve (see Figs.~\ref{res:bluedopp}
and \ref{res:reddopp}). These deviations are larger in the data
presented by Marsh (1999) when \gpc\ seemed to be in a different state
(as seen by comparing the average spectra). These deviations, probably
caused by phase-dependent visibility, are directly linked with flux
variability seen in the bright spot emission.

To study the flux variability in the emission lines as a function of
orbital phase we use a new extension to Doppler tomography. The method
is described in detail in Steeghs (2001, 2003), and splits the
reconstructed Doppler maps in an image describing the constant
contribution and two images describing the variable contribution. The
first of these two maps represents line flux varying with the cosine
of the phase, while the second represents flux varying with the sine
of the phase.  We are in particular interested in modelling the
anisotropic emission from the bright spot region in \gpc.
Since the strong central spike does contribute to the erratic flaring,
but does not modulate smoothly in intensity with orbital phase, we
masked out the low velocity regions of the line concerned. For all the
lines, the achieved $\chi^2$ values were significantly better using
the modulation mapping code, compared to the fits achieved by standard
Doppler tomography.
No significant residuals remained, indicating that most of the
anisotropies present were indeed modulated on the orbital period and
adequately prescribed by our modulation mapping code.

The persistent emission was discussed in Sect. 3.4 and revealed the
accretion disc, the stream-impact region and the central spike to be
the main emission line sources.  Modulation maps show that the disc
emission is only very weakly modulated, and that most of the
time-dependent flux originates in the bright spot region. We thus
limit our detailed discussion concerning the results from the
modulation mapping to the properties of the bright spot. In Fig.
\ref{res:modmaps}, the modulation maps of the bright spot region are
presented for \hee\ and \hef.
Considerable differences in the bright spot properties are found when
comparing lines. All lines show a considerable ($>$5 per cent)
time-dependent emission component from the bright-spot region reaching
for example an amplitude of 19 per cent in \hef. The location and
phasing of the variable bright-spot emission differs considerably from
line to line but is generally positioned at smaller $V_x$ and $V_y$
velocities compared to its persistent counterpart for a given line.
The lower $V_y$ velocities indicate that the modulated emission from
the bright spot region has a smaller contribution from material moving
with disc velocities, while the lower $V_x$ velocities indicate it is
further out in terms of radius. Thus, to first order, the bright spot
emission can be described by a steady component lying deeper into the
disc and showing some mixing with disc velocities, with additional
anisotropic emission localised along the incoming stream.
In some lines, apart from this compact modulated spot, there is also
extended modulated emission along the stream. In \hee, this stream
emission modulates at 3 per cent and peaks around phase 0.15, while in
\heb\ it dominates the variable emission and peaks close to phase
0.25.  In contrast, the \hea, \hef\ and \heh\ transitions do not show
such an extended modulated component.

The modulation maps thus highlight the complex morphology of the
bright spot emission in \gpc. Different lines are dominated by
different parts of the disc-stream interaction region. Unfortunately,
the uncertainty in both the mass ratio as well as the absolute phasing
does not allow us to compare the observed velocities to a ballistic
stream model in detail.

\section{UV spectra}

\begin{figure}
\begin{picture}(100,0)(-270,250)
\put(0,0){\includegraphics{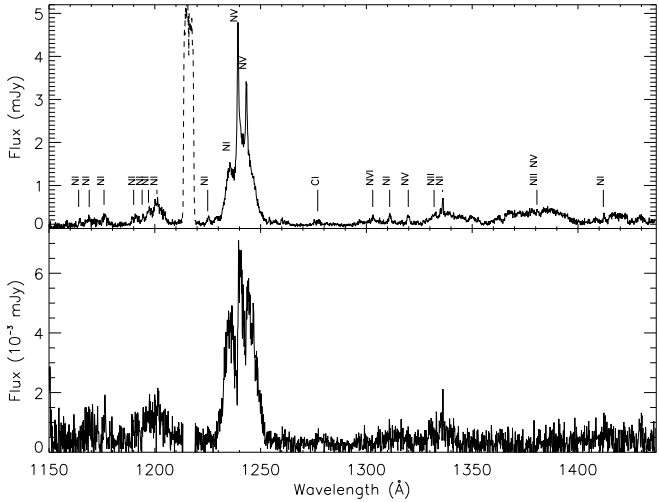}}
\noindent
\end{picture}
\vspace{68mm}
\caption{Top panel: average HST spectrum of \gpc\ in the UV. 
  Bottom panel: flare spectrum of \gpc\ in the same wavelength region.
  Most of the emission lines have been identified and labelled in the
  average spectrum.}
\label{res:hst1}
\end{figure}

\begin{figure}
\begin{picture}(100,0)(-270,250)
\put(0,0){\includegraphics{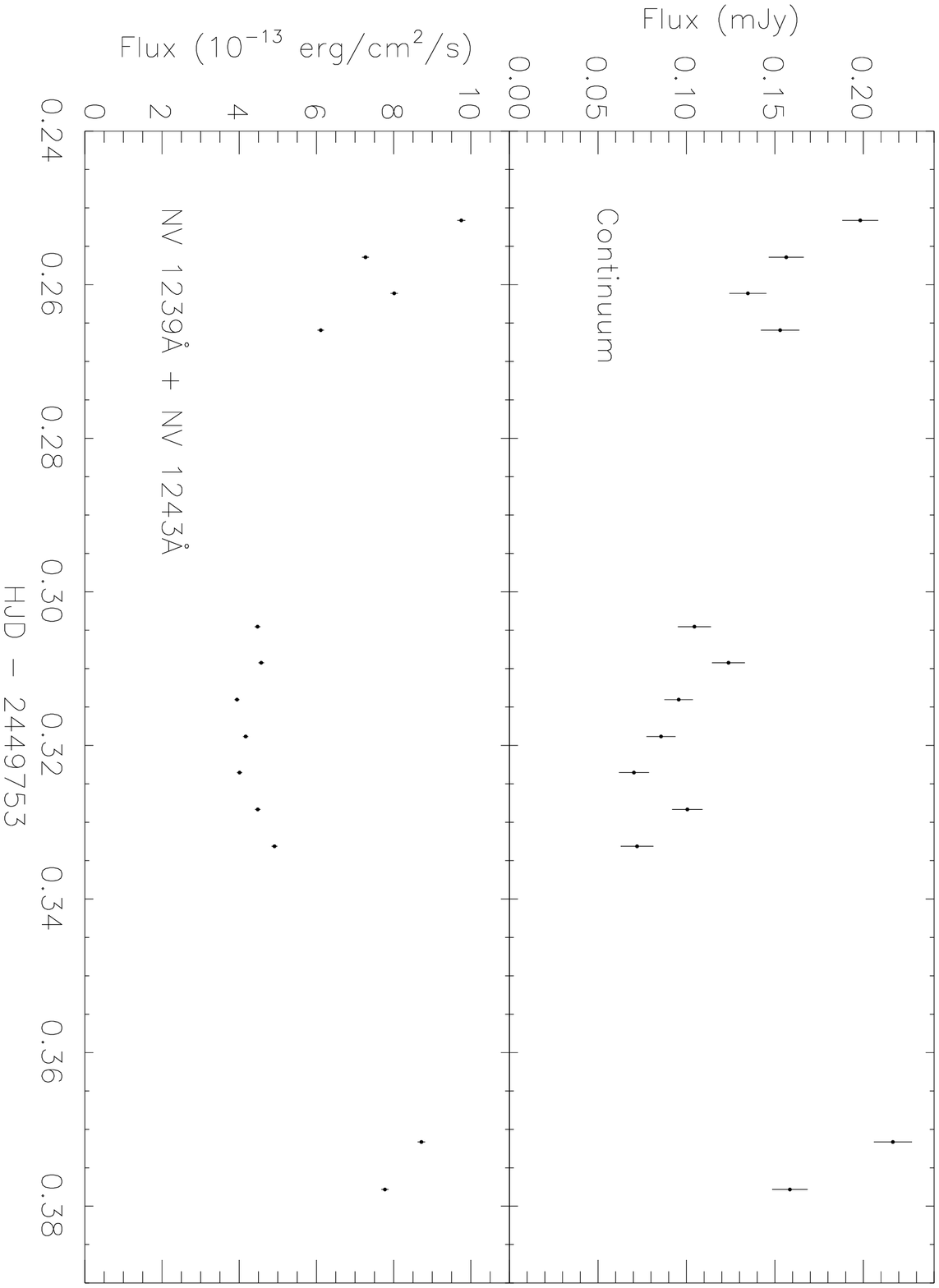}}
\noindent
\end{picture}
\vspace{68mm}
\caption{Lightcurves of the UV continuum (top) and the sum of \nva\
  and \nvb\ (bottom).}
\label{res:hstlc}
\end{figure}

\begin{figure}
\begin{picture}(100,0)(-270,250)
\put(0,0){\includegraphics{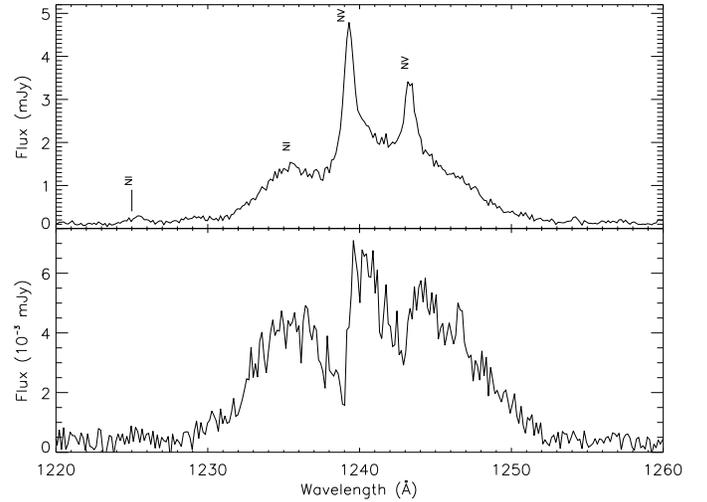}}
\noindent
\end{picture}
\vspace{68mm}
\caption{Same as in Fig.~\ref{res:hst1} but expanded on the N\,{\sc v}
  features.}
\label{res:hst2}
\end{figure}

The UV spectra are equivalent to those presented by Marsh et al.
(1995) but in this case the spectral resolution is higher and our
spectra do not suffer from offsets in the acquisition. The combination
of these two factors results in the blue wing of \nva\ not overlapping
with the geocoronal Lyman $\alpha$ emission (dashed section of the
spectrum in Fig.~\ref{res:hst1}), allowing us to see another N\,{\sc
  i} line bluewards of it.  The top panel of Fig.~\ref{res:hst1} shows
an average of the 13 spectra with the principal emission lines
labelled.  Most of the lines are nitrogen, {\sc i},{\sc ii}, and {\sc
  v} and possibly C\,{\sc i}. Of most interest because of their
strength are the N\,{\sc v} lines, \nva\ and \nvb. These two lines
seem to be composed of a central spike and broad wings like the
emission lines in the optical (see Fig.~\ref{res:hst2} for an expanded
view of these lines). Although we have labelled the feature on the
left of \nva\ as N\,{\sc i}, it is probably a combination of N\,{\sc
  i}\,$\lambda$1235\AA\ and the blue wing of \nva.

We computed Doppler maps (Marsh \&\ Horne 1988) of \nva\ and \nvb.
They show the low velocity emission associated with the central spike
in the lines and a ring-like structure situated at $\sim$1200\kmsec\ 
in the case of \nva\ and $\sim$1000\kmsec\ in the case of \nvb. These
structures are associated with the accretion disc.  The fact that the
velocities associated to the disc are larger in the UV
($\sim$1000\kmsec) than those seen for the optical emission lines
($\sim$800\kmsec) is due to the UV emission being restricted to the
inner regions of the disc. The low velocity emission also appears as a
ring around the zero velocity reaching of the order of 100\kmsec.
This is probably an artifact caused by the fact that we have assumed a
zero systemic velocity when computing the maps when the real systemic
velocity is not zero (see Marsh \&\ Horne 1988 for an explanation on
the origin of artifacts seen in Doppler maps). These maps were
obtained using only 13 spectra that, although they cover almost twice
the orbital period, give very low resolution maps, full of artifacts
in the form of radial features. This is why we do not display them.
There is no clear presence of emission at the bright spot position in
contrast with the data discussed by Marsh et al. (1995).

We notice that there are significant differences between each of the
13 spectra under study and conclude that there is a strong flare
component superposed to the average spectrum. We present the
lightcurves obtained for the continuum and the N\,{\sc v} lines in
Fig.~\ref{res:hstlc}. Although the number of spectra obtained is quite
small we can clearly see flux variations in the continuum and the
emission lines. The data is divided into three subsets with gaps of 7
and 8 hours respectively between the subsets. The continuum and line
flux variations follow the same trends, in contrast with the behaviour
observed for the optical blue data. We observe flux varibility in two
different timescales. Variations that lasts of the order of 20 min are
seen during the two first subsets of data (it is unclear whether these
are present during the third subset). We also see a difference in flux
between the three subsets, the first and third appearing brighter than
the second. In the same way discussed in Sect.~\ref{op:flares} we
calculate the flare component of the spectrum and display it in the
bottom panels of Figs.~\ref{res:hst1} and \ref{res:hst2}. The two
N\,{\sc v} lines as well as the fainter N\,{\sc i} lines contribute
significantly to the flare spectrum. It is not clear whether the
central spike is present in the flare spectrum. In contrast with Marsh
et al. (1995) we do not detect any new features in the flare spectrum.
We cannot confirm the presence of C\,{\sc iv}\,$\lambda$1550\AA\ as
our spectrum does not cover that wavelength region.  We do notice that
there is no emission in the Si\,{\sc iv}\,$\lambda$1400\AA\ doublet
region confirming that silicon is under-abundant in \gpc. If the
presence of nitrogen in \gpc\ is due to the action of the CNO process,
the only explanation for the high abundance of nitrogen and low
abundance of silicon in the system is that nitrogen-rich material was
transferred from the progenitor of the accreting white dwarf to its
companion during a common envelope phase.

\section{Conclusions}

We confirm the presence of the central spike in the flaring component
of the spectra which indicates that its origin is probably the
accreting star. We find that the central spike and the continuum vary
in a similar ways which results in the central spike not being present
for all lines in the flare spectrum obtained from the continuum
normalised data. This is not the case for the broad component of the
lines which indicates that they originate in different regions.

We detect the presence of a satellite peak bluewards of the central
spike for three of the emission lines (i.e. \hea, \heb\ and \hed) and
suggest that they are neutral helium forbidden transitions excited by
the presence of charged particles in the gas (the so called Stark
effect). If these satellite peaks are indeed the result of Stark
effect, we predict the presence of other lines associated with more
neutral helium forbidden transitions at 3705, 4009, 3820, 4144 and
specially 4026 \AA\ and encourage astronomers to observe \gpc\ as well
as CE315 at those wavelengths. We believe this could be the key to
understanding the origin of the until now elusive ``central spike''
and of the different spike redshifts seen for the different lines. The
comparative study of \gpc\ and CE315 will also contribute to the
understanding of these peculiar systems. We should keep in mind that
although this explanation is very attractive, the predicted and
measured separations between the forbidden and permitted components do
not match so there might be other effects contributing to the
behaviour of the central spike.

The Doppler maps computed from the spectra show the presence of an
elongated bright spot and a slightly elliptical accretion disc.  We
measure an eccentricity for the disc of 0.6. 

Modulation Doppler maps show that the line emission from the bright
spot is highly anisotropic. The anisotropic emission is formed at
slightly larger radii, whereas the persistent emission shows some
evidence of velocity mixing with the disc flow. 

From the UV spectra we conclude that there is no sign of the presence
of silicon in the system. The high nitrogen abundance combined with
the low silicon abundance can be explained if nitrogen-rich gas was
transferred from the progenitor of the white dwarf to the companion
during a common envelope phase.

\subsection*{Acknowledgements}

LMR was supported by a PPARC post-doctoral grant. DS acknowledges
support from a PPARC Fellowship and is currently supported by a
Smithsonian Astrophysical Observatory Clay Fellowship. EUS
acknowledges the support of a PPARC Gemini, Fundaci\'{o}n Andes
studentship for Ph.D studies at the University of Southampton. The
William Herschel and the Jacobus Kapteyn telescopes are operated on
the island of La Palma by the Isaac Newton Group in the Spanish
Observatorio del Roque de los Muchachos of the Instituto de
Astrof\'{\i}sica de Canarias. Based on observations made with the
NASA/ESA Hubble Space Telescope, obtained at the Space Telescope
Science Institute, wich is operated by the Association of Universities
for Research in Astronomy, Inc., under NASA contract NAS 5-26555.
These observations are associated with proposal \# 5689. The reduction
and analysis of the optical data were carried out on the Southampton
node of the STARLINK network.


\begin{thebibliography}{}
  
\bibitem[\protect\citename{Beauchamp \& Wesemael}1998]{bw98}
  Beauchamp\,A. \& Wesemael\,F., 1998, ApJ, 496, 395
\bibitem[\protect\citename{Beauchamp, Wesemael \&
  Bergeron}1997]{bwb97} Beauchamp\,A., Wesemael\,F. \& Bergeron\,P.,
  1997, ApJSS, 108, 559  
\bibitem[\protect\citenamee{Beauchamp, Wesemael, Bergeron \&
  Liebert}1995]{bwbl95} Beauchamp\,A., Wesemael\,F., Bergeron\,P. \&
  Liebert\,J., 1995, ApJ, 441, L85

\bibitem[\protect\citename{Cumming, Marcy \&\ Butler\,}1999]{cmb99}
  Cumming\,A., Marcy\,G.\,W. \& Butler\,R.\,P., 1999, ApJ, 526, 890

\bibitem{b01} Fitzgibbon, A., Pilu, M. \& Fischer, R., 1999, IEEE Trans.
PAMI, 21, 476
\bibitem{h69} Henon\,M., 1969, A\&A, 1, 223
\bibitem[\protect\citename{Lambert \&\ Slovak\,}1981]{ls81}
  Lambert\,D.\,L. \& Slovak\,M.\,H., 1981, PASP, 93, 477
\bibitem[\protect\citename{Marsh }1989]{h86} Marsh\,T.\,R., 1989,
  PASP, 101, 1032
\bibitem[\protect\citename{Marsh}1999]{m99} Marsh\,T.\,R., 1999,
  MNRAS, 304, 443
\bibitem[\protect\citename{Marsh, Horne \& Rosen}1991]{mhr91}
  Marsh\,T.\,R., Horne\,K. \& Rosen\,S., 1991, MNRAS, 366, 535
\bibitem[\protect\citename{Marsh}1995]{mwhl95} Marsh\,T.\,R.,
  Wood\,J.\,H., Horne\,K. \& Lambert\, D., 1995, MNRAS, 274, 452
\bibitem[\protect\citename{Nather, Robinson \& Stover}1981]{nrs81}
  Nather\,R., Robinson\,E. \& Stover\,R., 1981, ApJ, 244, 269
  
\bibitem[\protect\citename{Nelemans et al.}2001]{n01} Nelemans\,G.,
  Portegies Zwart\,S.\,F., Verbunt\,F. \& Yungelson\,L.\,R., 2001,
  A\&A, 368, 939

\bibitem[\protect\citename{Osaki}1974]{o74} Osaki\,Y., 1974, PASJ, 26, 429
\bibitem[\protect\citename{Osawa}1959]{o59} Osawa\,K., 1959, ApJ, 130, 159
  
\bibitem[\protect\citename{Podsiadlowski, Han \& Rappaport}2003]{phr}
  Podsiadlows\,Ph., Han\,Z. \& Rappaport\,S., in preparation
  (astro-ph/0109171)

\bibitem[\protect\citename{Ruiz} 2001]{r01} Ruiz\,M.\,T.,
  Rojo\,P.\,M., Garay\,G. \& Maza\,J., 2001, ApJ, 552, 679

\bibitem[\protect\citename{Savonije, de Kool \& van den Heuvel}1986]{skh86}
  Savonije\,G.\,J., de Kool\,M. \& van den Heuvel\,E.\,P.\,J., 1986,
  A\&A, 155, 51

\bibitem[\protect\citename{Scargle}1982]{s82} Scargle\,J.\,D., 1982,
  ApJ, 263, 835
\bibitem[\protect\citename{Schneider \& Young}1980]{sy80}
  Schneider\,D.\,P. \& Young\,P.\,J., 1980, ApJ, 238, 946
\bibitem[\protect\citename{Steeghs}2001]{s01} Steeghs\,D., 2001,
  Proceedings of the Astro-tomography workshop, Brussels, July 2000,
  eds. H.Boffin, D.Steeghs, J.Cuypers, Springer-Verlag Lecture Notes
  in Physics Series, Volume 573, page 45
\bibitem[\protect\citename{Steeghs}2003]{s03} Steeghs\,D., 2003,
  MNRAS, submitted
\bibitem[\protect\citename{Stone}1977]{s77} Stone\,R.\,P.\,S., 1977, AJ,
  218, 767
\bibitem[\protect\citename{Stover}1983]{s83} Stover\,R., 1983, PASP, 95, 18
\bibitem[\protect\citename{Tsugawa \& Osaki}1997]{to97} Tsugawa\,M. \&
  Osaki\,Y., 1997, PASJ, 49, 75
\bibitem[\protect\citename{Tutukov \&\ Federova\,}1989]{tf89}
  Tutukov\,A. \& Federova\, 1989, Sov. Astr., 33, 606
\bibitem[\protect\citename{Tutukov \&\ Yungelson}1979]{ty79} Tutukov\,A. \&
  Yungelson\,L., 1979, Acta Astron., 29, 665
\bibitem[\protect\citename{Tutukov \&\ Yungelson}1996]{ty96} Tutukov\,A. \&
  Yungelson\,L., 1996, MNRAS, 280, 1035
\bibitem[\protect\citename{van Teeseling \&\ Verbunt}1994]{vtv94} van
  Teeseling\, A \& Verbunt\,F., 1994, A\&A, 292, 519

\end{thebibliography}
\end{document}